\documentclass[aps,prb,longbibliography,twocolumn,groupedaddress]{revtex4-1}
\usepackage{graphicx}
\usepackage{sidecap}
\usepackage{amssymb}
\usepackage{amsmath}

\begin{document}


\title{Propagating Polaritons in III-Nitride Slab Waveguides}


\author{J. Ciers}
\email[]{joachim.ciers@epfl.ch}
\author{J. G. Roch}
\altaffiliation{Currently at Nano-Photonics Group, University of Basel, CH-4056 Basel, Switzerland}
\author{J.-F. Carlin}
\author{G. Jacopin}
\author{R. Butt\'e}
\author{N. Grandjean}

\affiliation{Institute of Physics, \'Ecole Polytechnique F\'ed\'erale de Lausanne (EPFL), CH-1015 Lausanne, Switzerland}


\begin{abstract}
We report on III-nitride waveguides with \textit{c}-plane GaN/AlGaN quantum wells in the strong light-matter coupling regime supporting propagating polaritons. They feature a normal mode splitting as large as 60 meV at low temperatures thanks to the large overlap between the optical mode and the active region, a polariton decay length up to 100 $\mu$m for photon-like polaritons and lifetime of 1-2 ps; with the latter values being essentially limited by residual absorption occurring in the waveguide. The fully lattice-matched nature of the structure allows for very low disorder and high in-plane homogeneity; an important asset for the realization of polaritonic integrated circuits that could support nonlinear polariton wavepackets up to room temperature thanks to the large exciton binding energy of 40 meV.

\end{abstract}


\maketitle

\section{Introduction}

Over the past 15 years, optical interconnects, which are essential building blocks for the realization of photonic platforms fully-integrated on a chip -- so-called photonic integrated circuits (PICs), have triggered a huge interest due to their promising potential in the field of information technology for the realization of small-footprint low-energy devices that may also offer a high clock rate and eventually support strong nonlinearity.\cite{Miller2009} 

Among the various systems at play, one such interesting platform deals with exciton-polaritons, hereafter called polaritons, in the waveguide geometry. Polaritons are quasiparticles that arise from the hybridization between a (confined) photon mode and a semiconductor exciton in the so-called strong coupling regime (SCR).\cite{Weisbuch1992} Their properties are therefore intermediate between those of photons and excitons and as such they benefit from the best of both worlds. Polaritons can propagate at near light speed and interact through their exciton fraction. This makes them ideal candidates for low-power active all-optical devices such as switches, optical transistors and logic gates.\cite{Liew2008,Amo2010} One of the inherent difficulties with all-optical devices is the lack of interaction between photons, which makes it nearly impossible to directly modify a light beam with a second one. To operate a nonlinear all-optical device made from traditional dielectric materials very high intensities are required, with typical values for the nonlinear refractive index on the order of $10^{-16}$ cm$^2$/W,\cite{boyd2008nonlinear} and present plasmonic alternatives are inherently lossy. \cite{Wurtz2008} A polariton-based device, however, combines a strong nonlinearity --- on the order of 10$^{-10}$ cm$^2$/W --- with a very fast response time on the order of a few picoseconds.\cite{Amo2010,Walker2015}

%

In order to reach the SCR between excitons and photons, and maximize the normal mode splitting $\Omega_{Rabi}$ between polariton eigenmodes, it is paramount to reduce any source of decoherence and maximize the exciton-photon coupling strength $g_0$. For a confined optical mode coupled to quantum well (QW) excitons, it can be expressed as \cite{Savona1995, Fox2006} 

\begin{equation}
g_0=\hbar\sqrt{\frac{e^2 N_{QW}^{eff} f_X}{2m_0 \varepsilon_0 n_{eff}^2 L_{eff}}},
\label{eq:g0}
\end{equation} 

where $\hbar$ is the reduced Planck constant, $e$ is the elementary charge, $N_{QW}^{eff}$ is the effective number of QWs (i.e. the nominal number of QWs weighted by the intensity of the optical mode profile) that are coupled to the optical mode, $f_X$ is the oscillator strength per unit area of the QW exciton, $m_0$ is the free electron mass, $\varepsilon_{0}$ is the permittivity of vacuum, $n_{eff}$ is the effective refractive index of the optical mode and $L_{eff}$ is the corresponding modal effective length.\cite{suppl} From this equation, we can see the interest in using QWs supporting large oscillator strength excitons and promoting structures with a large $\frac{N_{QW}^{eff}}{L_{eff}}$. In addition, in order to maintain the SCR up to room temperature or above, the QW excitons should be stable at these elevated temperatures. This requirement favors direct wide bandgap semiconductors which naturally have large exciton binding energies.\\

\begin{figure*}
\includegraphics[width=0.9\textwidth]{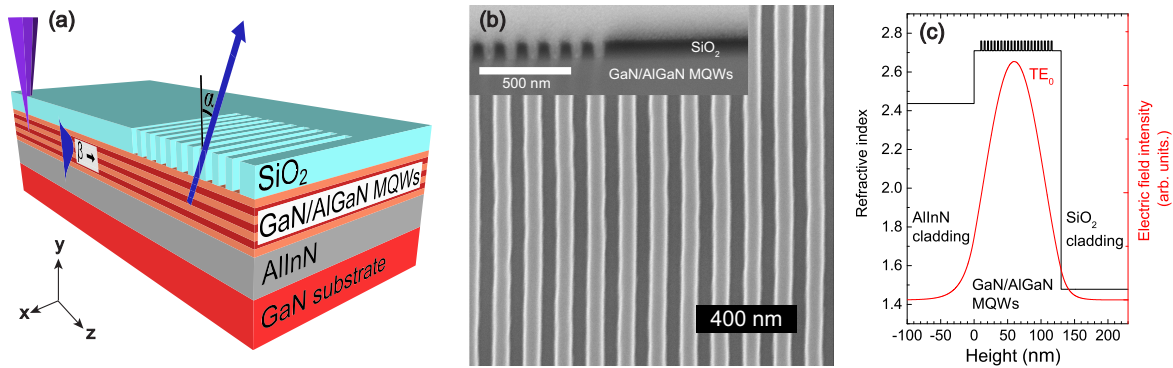}
\caption{(a) Sketch of the sample structure. (b) SEM of the fabricated gratings. Top view and cross-sectional view (inset). The cross-section was prepared by focused ion beam milling and the image is corrected for sample tilt. (c) Optical mode profile of the TE$_0$ mode supported by the present waveguide. This mode profile was calculated with a finite-difference time-domain mode solver. \cite{lumerical} The refractive index for GaN and AlGaN were taken from the work of Brunner \textit{et al.},\cite{Brunner1997} the refractive index of AlInN was taken from Butt\'e \textit{et al.}, \cite{Butte2005} and the refractive index of SiO$_2$ was taken from experimental data.}
\label{fig1}
\end{figure*}

As previously mentioned, the waveguide geometry is a promising platform for strong coupling applications.\cite{Ellenbogen2011a,Takeda2012,Walker2013, Solnyshkov2014a, Rosenberg2016} Indeed, the high in-plane group velocity on the order of $10^7$ m/s in waveguides makes them attractive for the achievement of fast PICs. Optical confinement provided by total internal reflection (TIR) results in smaller optical leakage and a small $L_{eff}$ value compared to distributed Bragg reflector (DBR) confinement traditionally used in planar microcavities (MCs). This is particularly the case for III-nitrides as the refractive index contrast between materials is low. The present III-nitride waveguide ---as detailed below--- supports a mode with an effective length of 175 nm, whereas a comparable III-nitride MC has an effective length of 590 nm.\cite{Christmann2008} A waveguide is also more robust against any deviations in layer thickness compared to MCs, since a guided mode does not depend on the cladding thickness and only slightly on the core dimensions, hence leading to higher fabrication yields. Additionally, the waveguide geometry can allow for easy potential landscaping by electrical gating \cite{Rosenberg2016} and electrical injection of carriers -- as is commonly done in edge-emitting laser diodes (LDs) -- as opposed to electrical injection through DBRs, which is a much more challenging task. As an illustration of the potential of strongly coupled waveguides, let us note that nonlinear effects relying on the formation of dark-bright spatio-temporal polariton soliton wavepackets with energies as low as 0.5 pJ has been recently reported in GaAs-based structures operating at 10 K.\cite{Walker2015}



In this work, we report on a III-nitride slab supporting waveguided polaritons originating from the hybridization of GaN/AlGaN multiple QW (MQW) excitons with the propagating TE$_0$ optical mode. Waveguided polaritons are monitored up to 100 K with a normal mode splitting of $\sim$60 meV. The guided polaritons propagate with a decay length of about 100 $\mu$m and have a lifetime of 1-2 ps. The sample structure is first discussed in Section \ref{sample}. The properties of the bare QW excitons and guided polaritons are then analyzed in Sections \ref{excitons} and \ref{polaritons}, respectively. The main conclusions and an outlook toward future work are given in Section \ref{conclusion}.


\section{\label{sample} Sample Structure}

The structure under investigation is sketched in Fig. \ref{fig1}(a) and consists of a 130 nm thick active region with 22 GaN/Al$_{0.1}$Ga$_{0.9}$N (1.5 nm/3.5 nm) QWs sandwiched between a 400 nm thick Al$_{0.83}$In$_{0.17}$N bottom cladding lattice-matched to GaN and a 100 nm thick SiO$_2$ top cladding. The bottom cladding and the active region were grown by metalorganic vapor phase epitaxy in an AIXTRON 200/4 RF-S reactor on a low dislocation density ($10^{6}$ cm$^{-2}$) freestanding (FS) \textit{c}-plane GaN substrate. High-quality AlInN layers have been demonstrated on these substrates\cite{Cosendey2011, Perillat-Merceroz2013} and have been successfully used for optical mode confinement in the waveguide region of visible III-nitride edge-emitting LDs.\cite{Castiglia2010} The AlInN bottom cladding layer contains seven 5 nm thick GaN interlayers positioned 50 nm apart in order to avoid kinetic roughening of the AlInN alloy. \cite{Perillat-Merceroz2013} Note that the effect of the quantum confined Stark effect (QCSE) on the exciton oscillator strength is negligible due to the reduced thickness (1.5 nm) of the present GaN/AlGaN QWs and the low Al content in the barriers.\cite{Grandjean1999} This is confirmed by the elevated value of 0.8 for the electron-hole overlap integral for these QWs, as calculated with the $k\cdot p$ formalism (see Section \ref{excitons} for further details). In addition, the use of an AlGaN barrier width of 3.5 nm avoids any coupling between adjacent wells. All aforementioned layer thicknesses are measured by high-resolution X-ray diffraction (HR-XRD). \cite{suppl} \\
 
A significant advantage of the polariton waveguide geometry over the more conventional planar MC design employing similar GaN/AlGaN QWs is that the bottom cladding can be grown lattice-matched to the FS-GaN substrate. Indeed in the MC case, a substantial Al concentration is required in the two quarterwave layers of the bottom UV DBR to avoid unwanted absorption at the GaN QW energy.\cite{Christmann2008} The resulting lattice-mismatch between the GaN template and the UV AlInN/AlGaN DBR requires the implementation of specific strain engineering solutions to avoid the formation of cracks\cite{Feltin2006} and leads to a higher density of defects, mainly threading dislocations, as there would be no additional benefit in using FS-GaN substrates in this latter case. The entire present structure is pseudomorphic to the FS-GaN substrate, as confirmed by HR-XRD reciprocal space mapping.\cite{suppl} \\

The SiO$_2$ top cladding was deposited by plasma-enhanced chemical vapor deposition on top of the active region. In order to outcouple the guided modes from the waveguide for subsequent analyses, a grating coupler was defined in the top cladding\cite{Taillaert2006} by electron beam lithography using a 100 keV Vistec EBPG5000 e-beam lithography system and ZEP520A positive resist, and inductively coupled plasma etching with CHF$_3$/SF$_6$ chemistry. A scanning electron micrograph (SEM) of the fabricated structure in top view and cross-section is shown in Fig. \ref{fig1}(b). The etched sidewalls form an angle of 3$^{\circ}$° with the vertical. The aspect ratio of the etched slits is 5:3. The gratings have a period $\Lambda$ of 125 nm with 50\% fill factor and span over a 100 $\times$ 100 $\mu$m$^{2}$ area. \\ 

We can relate the propagation constant $\beta$ of the guided mode to the emission angle $\alpha$ (shown in Fig. \ref{fig1}(a)) from the grating output coupler by the relation 
\begin{equation}
k_{z,air}=\frac{\omega}{c}\sin \alpha = \beta - \frac{2\pi q}{\Lambda},
\label{gratingk}
\end{equation}
where $q \in \mathbb{Z}$ is the diffraction order. We use a first-order grating to maximize the outcoupled light intensity. The grating period is chosen such that the central in-plane wavevector of interest (50 $\mu$m$^{-1}$) is emitted perpendicular to the sample.\\

A slab waveguide can support both transverse electric (TE, with the electric field along \textit{x}) and transverse magnetic (TM, with the magnetic field along \textit{x} and the electric field in the \textit{yz}-plane) guided modes. We define \textit{z} as the propagation direction of the guided modes, \textit{y} the growth direction of the sample and \textit{x} the direction perpendicular to both of these, as shown in Fig. \ref{fig1}(a). The thickness of the waveguide core was chosen such that only the TE$_0$ and TM$_0$ modes are supported and all higher order modes are cut off. This prevents the excitons from coupling to multiple modes, which would act as a lossy channel for the photogenerated carriers. The X$_A$ exciton couples exclusively, as X$_B$ mostly does, to the in-plane electric field as deduced from $k\cdot p$ calculations, which favors the TE mode over the TM one. The overlap integral of the $E_z$ field of the TM$_0$ mode with the active region is almost two orders of magnitude smaller than the overlap integral of the TE$_0$ $E_x$ field with the active region. Therefore, we will only consider the TE$_0$ mode in the remaining part of this work.\\

The present structure was optimized to maximize the light-matter coupling strength $g_0$ between the guided photons and MQW excitons. The high number of QWs (22) combined with the large overlap between the optical mode and the QWs -- the TE$_0$ mode has no nodes and the mode intensity quickly decreases outside the active region, as can be seen in Fig. \ref{fig1}(c) -- and the large oscillator strength of the QWs result in a high $g_0$ value of 30 meV.

\section{\label{excitons} Bare Quantum Well Properties}
\begin{figure}{}
\includegraphics[width=0.45\textwidth]{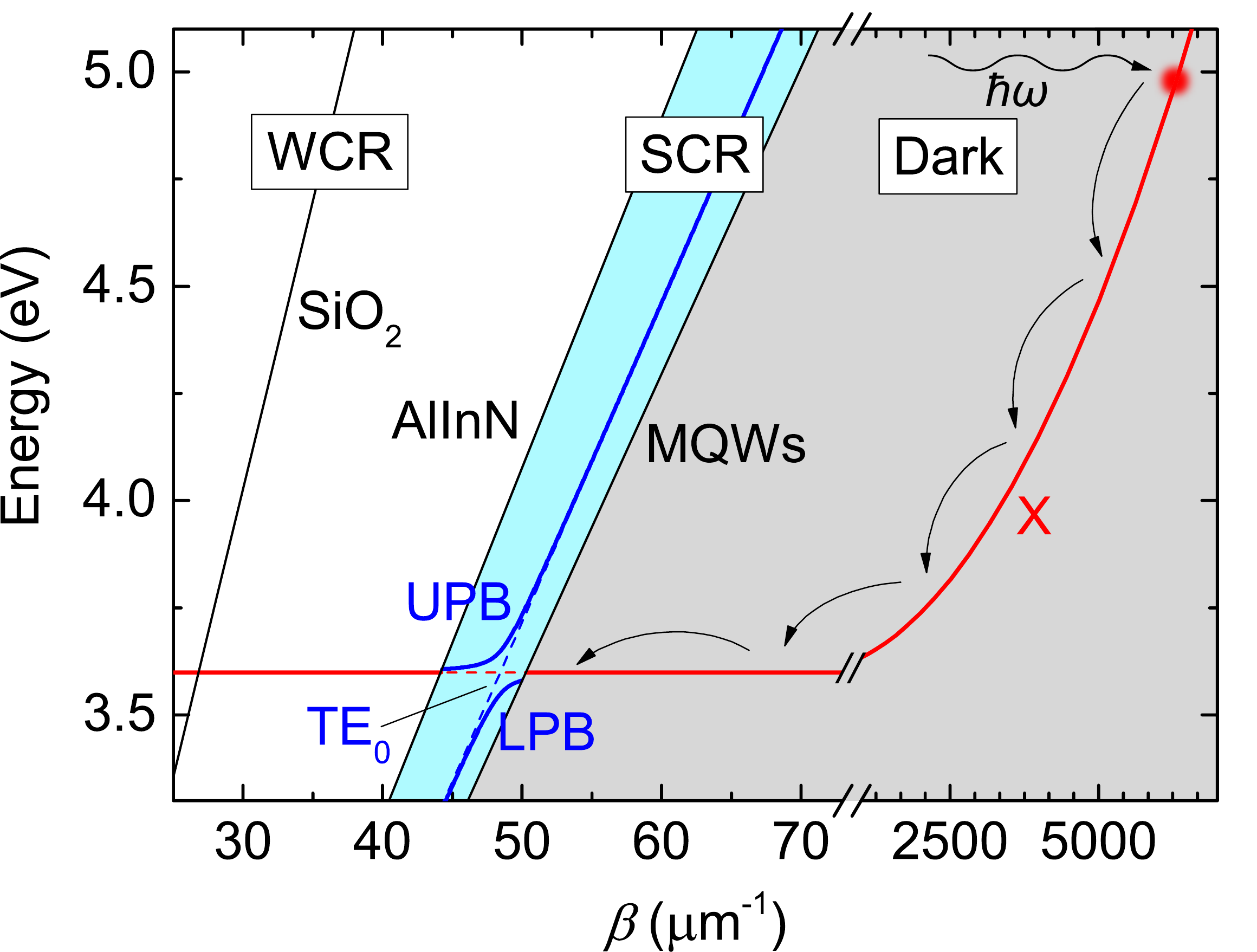}
\caption{Eigenmode dispersion of the present structure for polarization along $x$. The light cones of the multiple quantum wells (MQWs), bottom cladding (AlInN) and top cladding (SiO$_2$) are represented in black. They mark the transitions between the different light-matter coupling regimes. The dark regime occurs outside the active region light cone (marked in grey) where photons do not exist. Between the active region and the bottom cladding light cone, the exciton (X, red) and TE$_0$ guided mode are in the strong coupling regime (SCR, blue) and form an upper (UPB) and a lower polariton branch (LPB). Within the bottom cladding light cone, photons are poorly confined and couple weakly with the excitons (WCR, white). For simplicity, refractive index dispersion was neglected in this plot. The considered optical refractive index values are those expected at 3.6 eV (2.756 for the active region,\cite{Brunner1997} 2.42 for the bottom cladding \cite{Butte2005} and 1.47 for the top cladding). The exciton dispersion is calculated within the effective mass approximation.}
\label{couplingregimes}
\end{figure}

In the present structure, three different light-matter coupling regimes can potentially coexist in distinct zones of the energy vs. $\beta$ diagram as shown in Fig. \ref{couplingregimes}. Photons with an in-plane wavevector outside the active region light cone do not exist; hence explaining the terminology dark excitons as the radiative recombination of excitons in this region is forbidden. In the $k$-space region between the bottom cladding and the active region light cone, photons are tightly confined in the waveguide by TIR and, as will be shown in Section IV, they hybridize with the X$_{A}$ excitons. Inside the bottom cladding light cone, TIR is lost at the bottom interface and the active region becomes a lossy resonator with broad optical modes. Therefore excitons and photons are weakly coupled in this region. Inside the top cladding light cone, TIR is also suppressed at the top interface and the structure forms a low quality factor ($Q$) Fabry-Perot resonator. As a result, we simultaneously have three populations of excitations in the sample under cw non-resonant excitation: dark excitons, waveguided polaritons, and weakly coupled excitons. The latter population allows us to probe the bare excitonic properties inside the air light cone using conventional optical spectroscopy techniques.

\begin{SCfigure*}{}
\includegraphics[width=0.75\textwidth]{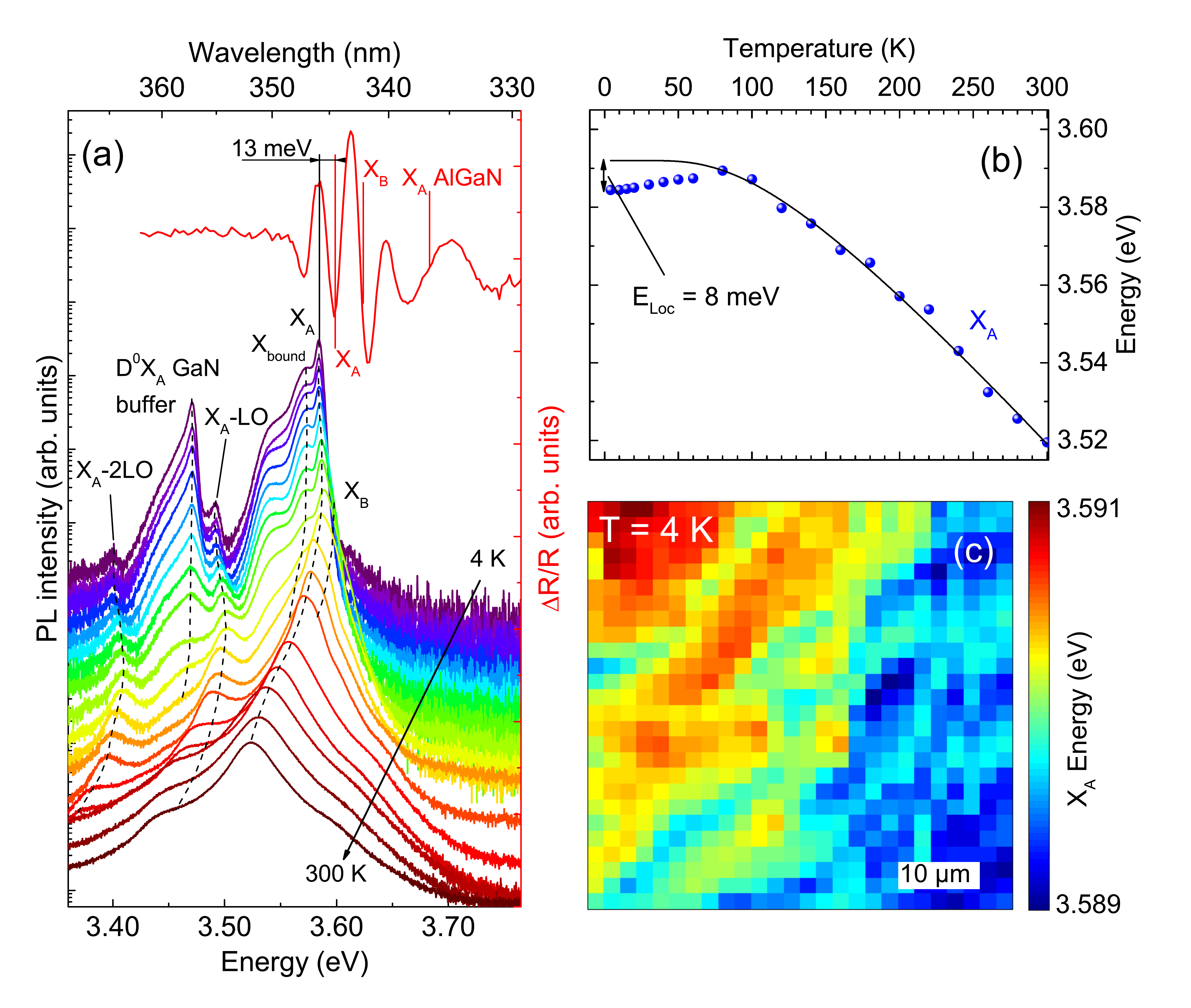}
\caption{(a) Low temperature ($T$ = 10 K) PR (red) and temperature series between 4 and 300 K of $\mu$-PL spectra taken at approximately the same location showing a Stokes shift of 13 meV. The spectra are vertically shifted for clarity. (b) Temperature dependence of the X$_A$ energy for the measurements shown in (a), together with a fit according to Eq. \ref{vinamodel} (black). (c) $\mu$-PL mapping of the X$_A$ emission energy measured at 4 K. The observed standard deviation is as low as 0.42 meV over a 50 $\times$ 50 $\mu$m$^{2}$ area.}
\label{PL}
\end{SCfigure*}

We calculated the confined MQW electron and hole energy levels at 0 K using the $k\cdot p$ formalism\cite{suppl,Chuang1996} adapted for the analysis of strained MQW structures. Due to the geometrical effect in the MQW, the value of the electric field found by setting the potential difference between the extremities of the active region to zero, amounts to 220 kV/cm in the AlGaN barriers and -625 kV/cm in the GaN wells, respectively.\cite{Bernardini1999} An excitonic energy $E_{X_{A}}=$ 3.566 eV and $E_{X_{B}}=$ 3.576 eV is obtained for X$_A$ for X$_B$, respectively, assuming an exciton binding energy of 40 meV in both cases. The latter was deduced from the variational approach developed by Leavitt and Little.\cite{Leavitt1990} The relative oscillator strength of the X$_A$ (0.5 for light polarized along both $x$ and $z$, 0 along $y$) and X$_B$ excitons (0.495 for light polarized along both $x$ and $z$, 0.01 along $y$) shows an exclusive coupling for X$_A$ and a heavily preferential coupling for X$_B$ to the in-plane electric field. Hence it explains the poor excitonic coupling to the TM modes supported by the waveguide.\\

The bare exciton properties were experimentally investigated by photoreflectance (PR) and micro-photoluminescence ($\mu$-PL) spectroscopy.{\cite{suppl}} Clear signatures can be observed in the low temperature PR spectrum (shown in Fig. {\ref{PL}}(a)) for the X$_A$ and X$_B$ QW excitons as well as the X$_A$ barrier exciton. No signature of the underlying FS-GaN substrate was observed in PR as with this technique the modulation induced by photogenerated carriers essentially occurs in the topmost layers. The critical points were fitted using the approach introduced by Aspnes.\cite{Aspnes1973,Glauser2014} The comparison between low-temperature PL and PR spectra (Fig. {\ref{PL}(a)}) reveals a Stokes shift of $13$ meV, which is similar to that previously reported for equivalent GaN/AlGaN MQW samples.{\cite{Glauser2014}} Note that since the PL and PR measurements were conducted with the sample in different cryostats, there is an uncertainty of $\sim$500 $\mu$m ---comparable to the PR spot size--- on the relative sample location probed in both experiments.  On the low-temperature PL spectra, we can clearly identify the X$_A$ MQW exciton together with its first and second longitudinal optic (LO) phonon replicas, each separated by $\sim$92 meV, the accepted LO-phonon energy for GaN. An inhomogeneous broadening of 8 meV is measured at 4 K for the X$_A$ peak. This very low value for a sample that consists of 22 QWs indicates the high quality of the latter, which compares favorably with state-of-the-art GaN/AlGaN QW samples.{\cite{Feltin2007,Stokker-Cheregi2008,Glauser2014}} On the low-energy side of the X$_A$ exciton, we observe a shoulder which is likely due to a bound excitonic states, X$_{bound}$. It could originate from the incorporation of acceptor-like impurities such as carbon during the low-temperature growth of the MQWs.{\cite{Cingolani1997,Weisbuch2016}} At low temperatures, a third peak is present at $\sim$3.55 eV. This energy corresponds to the maximum of the polariton emission ---see below--- and as such, this peak could be the signature of scattered lower polaritons. \\

The temperature dependence of the X$_A$ transition can be well described by the expression:{\cite{Vina1984}}

\begin{equation}
E_{X_A}(T)=E_{X_A}(0)-\frac{2 \alpha_{B}}{exp(\Theta_B/T)-1},
\label{vinamodel}
\end{equation} 

where $E_{X_A}(0)$ is the expected energy of the free $X_{A}$ exciton transition at 0 K and the second term is the Bose-Einstein occupation factor for phonons where $\alpha_{B}$ is an electron-phonon coupling constant and $\Theta_{B}$ is an average phonon temperature. This model is notably more accurate than Varshni's empirical law at low temperature.{\cite{Brunner1997}} Upon fitting the measured PL emission energies to Eq. {\ref{vinamodel}} (Fig. {\ref{PL}}(b)), we find a localization energy of 8 meV, together with values of 70 meV and 322 K for $\alpha_{B}$ and $\Theta_B$, respectively. These values for the latter two parameters are in good agreement with those reported by Brunner \textit{et} \textit{al.} for bulk (Al)GaN layers.{\cite{Brunner1997}}\\

The energy of the X$_A$ free exciton deduced fom PR measurements (3.599 eV at 10 K) and Eq. {\ref{vinamodel}}  (3.592 eV at 4 K) is slightly higher than that originally expected from $k\cdot p$ calculations (3.566 eV). This could be accounted for assuming a smaller value of the built-in electric field in the MQWs and slightly different in-plane carrier masses.\\

The in-plane homogeneity of the sample was checked through $\mu$-PL mapping performed at 4 K (Fig. \ref{PL}(c)), which leads to a standard deviation $\sigma$ as low as 0.42 meV for the X$_A$ energy over a 50 $\times$ 50 $\mu$m$^{2}$ area. This shows the high degree of homogeneity of the sample, which is an important asset for the control of waveguided polaritons and the potential realization of PICs relying on such an approach. The uniformity of those waveguide samples can likely be ascribed to the growth performed on low defect density FS-GaN substrate, as these latter samples are expected to be much less affected by in-plane disorder than their strongly coupled counterparts grown on \textit{c}-plane sapphire substrate. \cite{Feltin2007, Rossbach2013}  \\


To summarize this section, the low values of the inhomogeneous PL broadening (8 meV), localization energy (8 meV), Stokes shift (13 meV) and in-plane inhomogeneity ($\sigma$ = 0.42 meV) highlight the high quality and the homogeneity of the present sample, especially when taking into account the large number of QWs. \\

\section{\label{polaritons} Guided Polaritons}

The waveguide dispersion curve was measured for various propagation distances by moving the excitation spot away from the grating outcoupler between 4 and 100 K. The measured signal intensity is relatively weak for two reasons. First, only excitons lying outside the cladding light cone are expected to form polaritons, as shown in Fig. \ref{couplingregimes}, hence the excitonic fraction that relaxes into this light cone does not contribute to the signal. Second, as the slab waveguide is radially symmetric, waveguided polaritons, if present, will propagate isotropically from the excitation spot. Therefore, only polaritons within the angle covered by the collection zone can contribute to the signal and its intensity will approximately decrease as the reciprocal of the propagation distance $r$. This approximation holds well if the size of the collection region is substantially smaller than the propagation distance. Here, the collection region is 19 $\mu$m and the propagation distances used in the experiments range between 30 and 200 $\mu$m. Additionally, the internal quantum efficiency of the active medium decreases with increasing temperature, which further reduces the signal intensity at higher temperatures. Above 100 K, the signal intensity became too weak to perform any reliable measurements. Note however, that this does not imply that the SCR could not be maintained above 100 K.\\

\begin{figure*}{}
\includegraphics[width=0.80\textwidth]{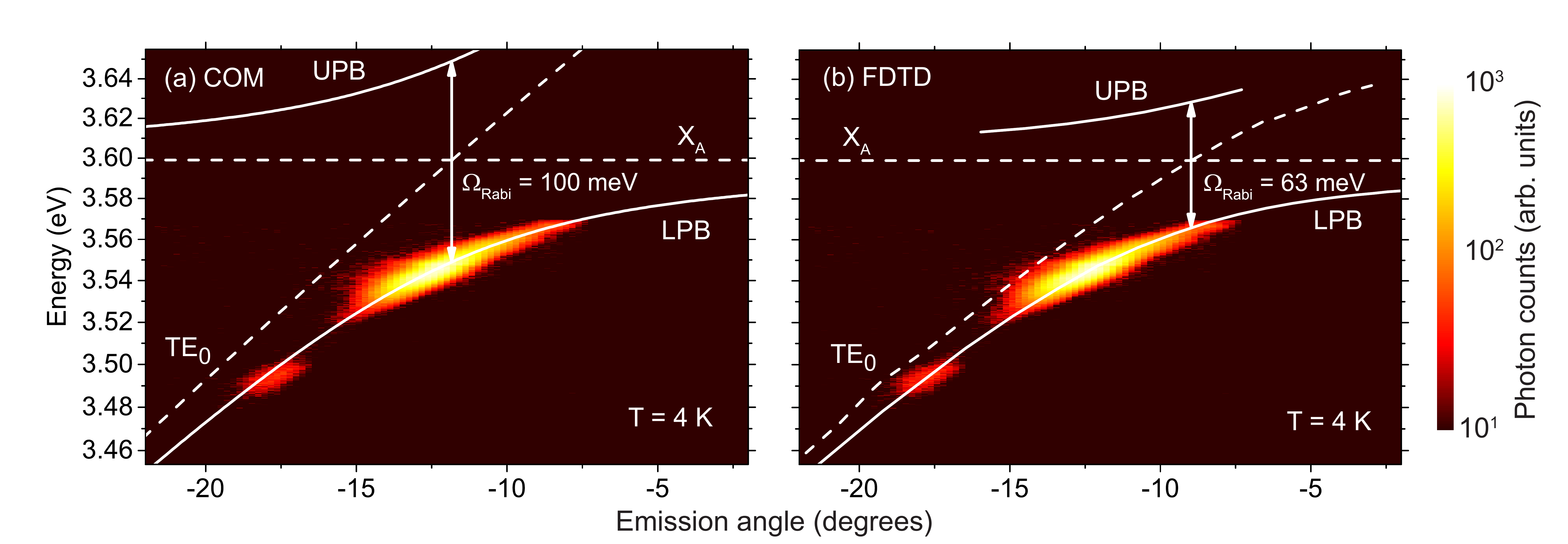}
\caption{Measured polariton dispersion for a propagation distance of 55  $\mu$m measured at $T$ = 4 K with an excitation power density of 640 W/cm$^2$. The background signal from the bare excitons was subtracted for clarity. \cite{suppl} (a) A conventional COM is fitted to the data and the resulting UPB and LPB are shown in white, together with the uncoupled exciton (X$_A$) and the TE$_0$ mode. This model leads to a normal mode splitting of 100 meV. (b) Full 2D-FDTD mode calculations were performed for various values of the exciton oscillator strength. A good correspondence with the measurement was found for $f_X = 1.1 \times 10^{13}$ cm$^{-2}$ (shown in white). We find a normal mode splitting of 63 meV using this method. The same experimental data are shown in both (a) and (b).}
\label{dispersion}
\end{figure*}

An example of a dispersion curve measured at $T$ = 4 K is shown in Fig. \ref{dispersion} where the background signal coming from bare excitons was removed for clarity.\cite{suppl} We clearly see a pronounced curvature in the measured signal, which is ascribed to the avoided crossing of lower polaritons with the uncoupled free X$_A$ transition.\cite{XBnote} We compared the measured signal at 4 K to both a simple coupled oscillator model (COM) and full two-dimensional finite-difference time-domain (2D-FDTD) calculations.{\cite{suppl}} The COM uses a constant value of $n_{eff}$, and therefore a linear dispersion for the uncoupled TE$_0$ waveguide mode. By fitting this model (the results are shown in Fig. {\ref{dispersion}}(a)), we find a normal mode splitting of 100 meV for this measurement. Application of Eq. {\ref{eq:g0}} results in an oscillator strength of $3.0 \times 10^{13}$ cm$^{-2}$ for the QW excitons. The 2D-FDTD calculations on the other hand take the material refractive index dispersion into account. The resulting uncoupled TE$_0$ mode is curved due to the increase in refractive index of the QWs and barriers near the band edge, located at 3.64 and 3.71 eV, respectively, as shown in Fig. {\ref{dispersion}}(b). The 2D-FDTD calculations were performed for various values of $f_X$, and a good fit to the data was found for $f_X=1.1\times 10^{13}$ cm$^{-2}$. The corresponding normal mode splitting is 63 meV. The discrepancy in the values of $f_X$ and $\Omega_{Rabi}$ between the COM and the 2D-FDTD calculations is mainly due to the neglected refractive index dispersion in the former case. Since the uncoupled mode is bent in the same direction as the anticrossing by the increasing refractive index toward the band edge, a smaller oscillator strength is required to reproduce the measured dispersion. A COM featuring a constant effective refractive index therefore systematically overestimates the exciton oscillator strength and the normal mode splitting. This comparison shows the importance of the effect of the refractive index dispersion in the analysis of photonic structures operating near the band edge. Note that a value of $f_X=2.1\times 10^{13}$ cm$^{-2}$ was previously reported for similar QWs embedded in a strongly coupled MC using transfer matrix simulations accounting for dispersion.{\cite{Glauser2014}} Let us point out that the observed curvature in the dispersion cannot be explained by the increase in the refractive index of the active region near the band gap as can be seen in Fig. {\ref{dispersion}}(b), which confirms that the present structure is operating in the SCR. \\

We do not observe the highly excitonic tail of the LPB, since the polaritons relax to lower energy states and because the outcoupling rate is proportional to their photon fraction. The photonic tail of the LPB is hardly visible in the measured polariton dispersions, most likely due to the relaxation bottleneck of highly photonic polaritons.{\cite{Tassone1997}} Note that there is a second maximum in the polariton emission intensity at $\sim$92 meV below the localized X$_A$ emission energy. This could be explained by the LO-phonon assisted relaxation of excitons to the LPB as already reported in the planar MC case.{\cite{Corfdir2012}}\\

By comparison with 2D-FDTD calculations, we find a normal mode splitting of 63 meV for the curve measured at 4 K (Fig. {\ref{dispersion}}(b)), and an average splitting of 60 meV between 4 and 100 K. Such a value has to be compared to the 56-60 meV reported by Christmann \textit{et} \textit{al}. for a III-nitride planar MC containing 67 GaN/AlGaN QWs.{\cite{Christmann2008a}} The similar value of $\Omega_{Rabi}$ recorded for the two geometries is the direct manifestation of the increased $\frac{N_{QW}^{eff}}{L_{eff}}$ value in waveguide structures.\\

The absence of the UPB in the PL spectra is a well-known feature of wide band gap systems,\cite{Christmann2008, Guillet2011} which is mainly due to absorption occurring above the MQW band gap and the large $\Omega_{Rabi}$ value, which hinders the thermal promotion of polaritons to the UPB.\cite{Tassone1997} However, let us note that even in GaAs-based polariton waveguides, the UPB luminescence is usually rather weak or even absent.\cite{Walker2013,Tinkler2015,Rosenberg2016}\\

We did not observe any renormalization effects, i.e., any decrease in the normal mode splitting, with increasing pumping power up to 6 kW/cm$^2$. This is because we cannot measure any dispersion relation for propagation distances shorter than 20 $\mu$m. For these short distances, the outcoupled guided light is not distinguishable from the very intense direct PL signal originating from the MQWs inside the air light cone. Therefore, we essentially probed regions where the reduction in the polariton density coming from their radial outspread is significant and renormalization should not be expected.{\cite{Rossbach2013,Sturm2014}}\\

\begin{figure}{}
\includegraphics[width=0.5\textwidth]{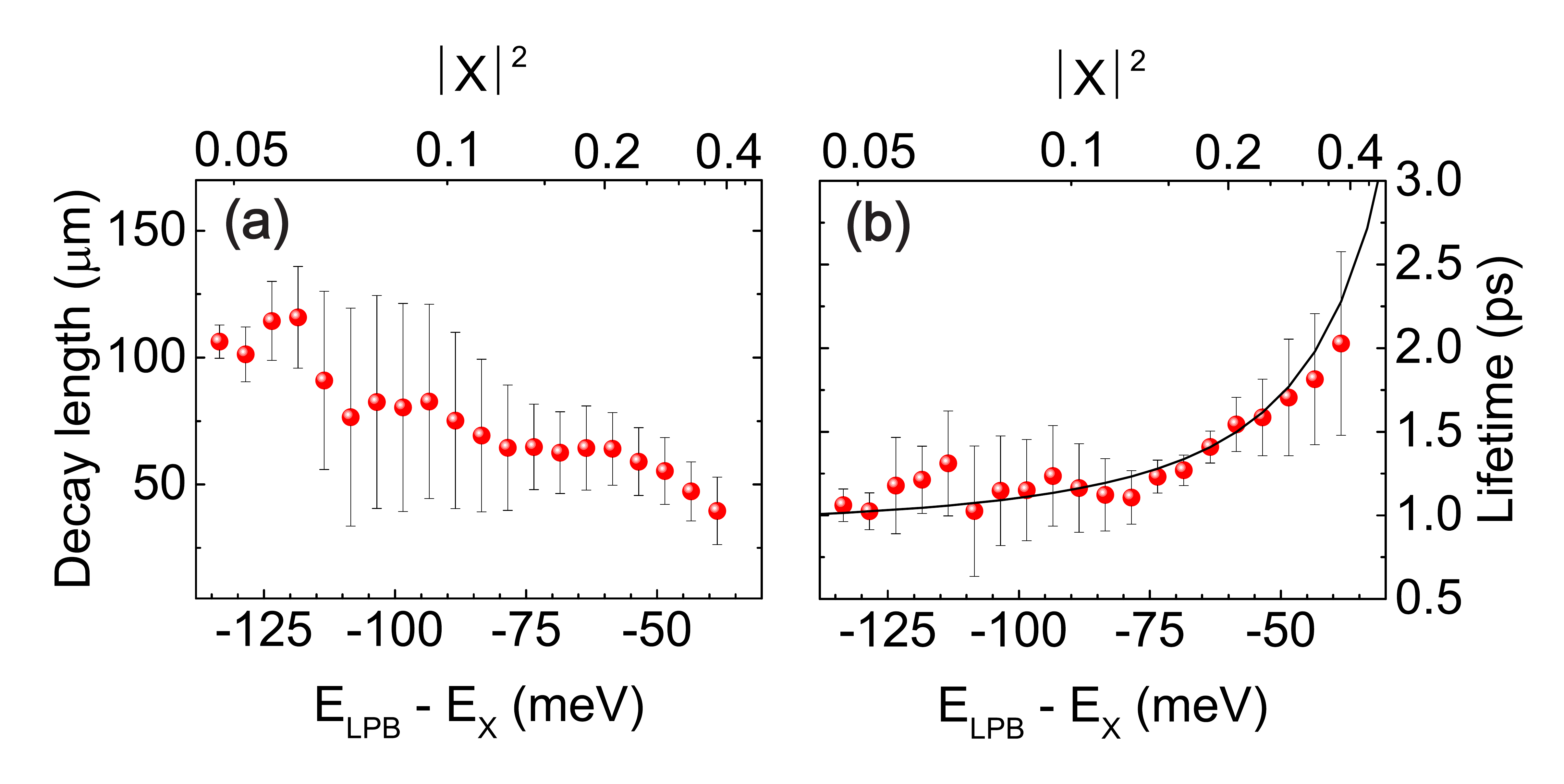}
\caption{Low temperature decay of waveguided polaritons. (a) Polariton decay length as a function of polariton energy. The mean value and error bars were deduced from a set of several measurements taken between 4 and 30 K. No substantial difference in the lifetime and the decay length was observed within this temperature range. (b) Lifetime of the guided polaritons. Measured data (red dots) and fit to Eq. \ref{eqlifetime} (black line). The top axis in both graphs represents the excitonic fraction of polaritons at the corresponding LPB energy.}
\label{Decaylengthlifetime}
\end{figure}

In addition to the above-mentioned temperature dependence, the observed decrease in the signal intensity with increasing propagation distance is not only due to the radial spread-out of waveguided polaritons from the excitation spot -- which leads to a decrease in the polariton density overlapping with the collection area as $r^{-1}$--  but also to polaritonic decay. Indeed, as waveguided polaritons propagate, they experience an intrinsic decay following an exponential Beer-Lambert law, mainly due to photonic losses for the present exciton-photon detunings ($\delta$). In order to determine their intrinsic decay length -- defined as the distance over which the integrated polariton PL intensity at the excitation spot is reduced by a factor $e$ -- and hence their lifetime, the measured integrated PL intensity is multiplied by the propagation distance which is fitted to an exponential decay. The obtained decay length as a function of polariton energy is shown in Fig. \ref{Decaylengthlifetime}(a). Polaritons with a large photonic fraction in excess of 0.94 are characterized by a decay length larger than 100 $\mu$m, which decreases as the excitonic fraction becomes larger. To further support this analysis, the absorption coefficient in the waveguide was determined by the variable stripe length method to be 60 cm$^{-1}$ at around 200 meV below the free X$_A$ energy,\cite{suppl} which corresponds to a decay length of 167 $\mu$m. This puts an upper limit to the propagation length of photons and highly photonic polaritons in the present sample and is mainly limited by residual absorption from the QWs. We can expect the absorption to be higher ---and the corresponding decay length lower--- for higher energies. Since the value of 167 $\mu$m only slightly exceeds the measured polariton decay lengths at energies between 120 and 140 meV below X$_A$, absorption is most likely the limiting factor of polariton propagation.\\

Another essential figure of merit to qualify the present structure is the polariton lifetime $\tau_{pol}(\beta)$. The latter is given by 

\begin{equation}
\frac{1}{\tau_{pol}(\beta)}=\frac{\vert P(\beta) \vert^2}{\tau_{p}}+\frac{\vert X(\beta) \vert^2}{\tau_{X}},
\label{eqlifetime}
\end{equation} 

where $\tau_{p}$ and $\tau_{X}$ are the photon and exciton lifetime, respectively, $P$ and $X$ are the usual Hopfield coefficients and $\vert P(\beta) \vert^2$ and $\vert X(\beta) \vert^2$  give the photon and exciton fraction of the polaritons, respectively.{\cite{Hopfield1958}} The polariton lifetime $\tau_{pol}(\beta)$ was determined by dividing the measured polariton decay length by the group velocity of the LPB, defined as $\frac{\partial \omega}{\partial \beta}$, (Fig. \ref{Decaylengthlifetime}(b)). Values ranging between 1 and 2 ps were derived for polaritons with an exciton fraction between 0.05 and 0.4. The free X$_A$ lifetime was determined by time-resolved PL to be 325 ps at $T = 10$ K. A fit from Eq. \ref{eqlifetime} to the data leads to a value of 0.9 ps for $\tau_{p}$, which corresponds to a photonic linewidth $\gamma_P = 0.73$ meV and an optical quality factor of $4.9\times10^{3}$. This photonic lifetime is about a factor of 5 larger than that reported for comparable III-nitride planar MCs,\cite{Levrat2010} essentially due to the improved confinement provided by TIR. Note that we do not directly observe the photonic and excitonic linewidths stated above in the measured dispersion signal for highly photonic and excitonic polaritons, respectively. In addition to this broadening along the energy axis, there is indeed an extra source of broadening in the emission angle due to fabrication inhomogeneities in the grating coupler. For highly photonic polaritons, the energy linewidth becomes relatively small and the observed signal broadening is of purely angular origin. In this limit, we observe an angular full width at half maximum linewidth (FWHM) $\Delta \alpha$ of 1.2$^{\circ}$, which corresponds to a FWHM wavevector linewidth $\Delta k_z$ of 0.34 $\mu$m$^{-1}$ and a FWHM error on the grating period $\Delta \Lambda = 0.8$ nm through Eq. {\ref{gratingk}}. The normal mode splitting of 60 meV corresponds to a Rabi period, i.e., the period of the coherent oscillations between the photon and exciton fraction of the polariton, as short as 69 fs. In this respect, we point out that the present sample is characterized by a large polariton lifetime to Rabi period ratio. When comparing the present values to those obtained in GaAs-based waveguides,\cite{Walker2013} the shorter decay lengths and lifetimes in III-nitride structures are likely due to the combination of an enhanced sensitivity to photonic disorder of short wavelength systems together with the increased QW absorption below the band gap and the high number of QWs. If we compare the obtained values for the decay length to the ballistic condensate propagation on the order of 10 $\mu$m reported by Hahe \textit{et} \textit{al.} in a ZnO planar MC,{\cite{Hahe2015}} the much larger decay length in the present case can be well accounted for by TIR confinement and the larger polariton propagation velocity.\\

\begin{figure}{}
\includegraphics[width=0.45\textwidth]{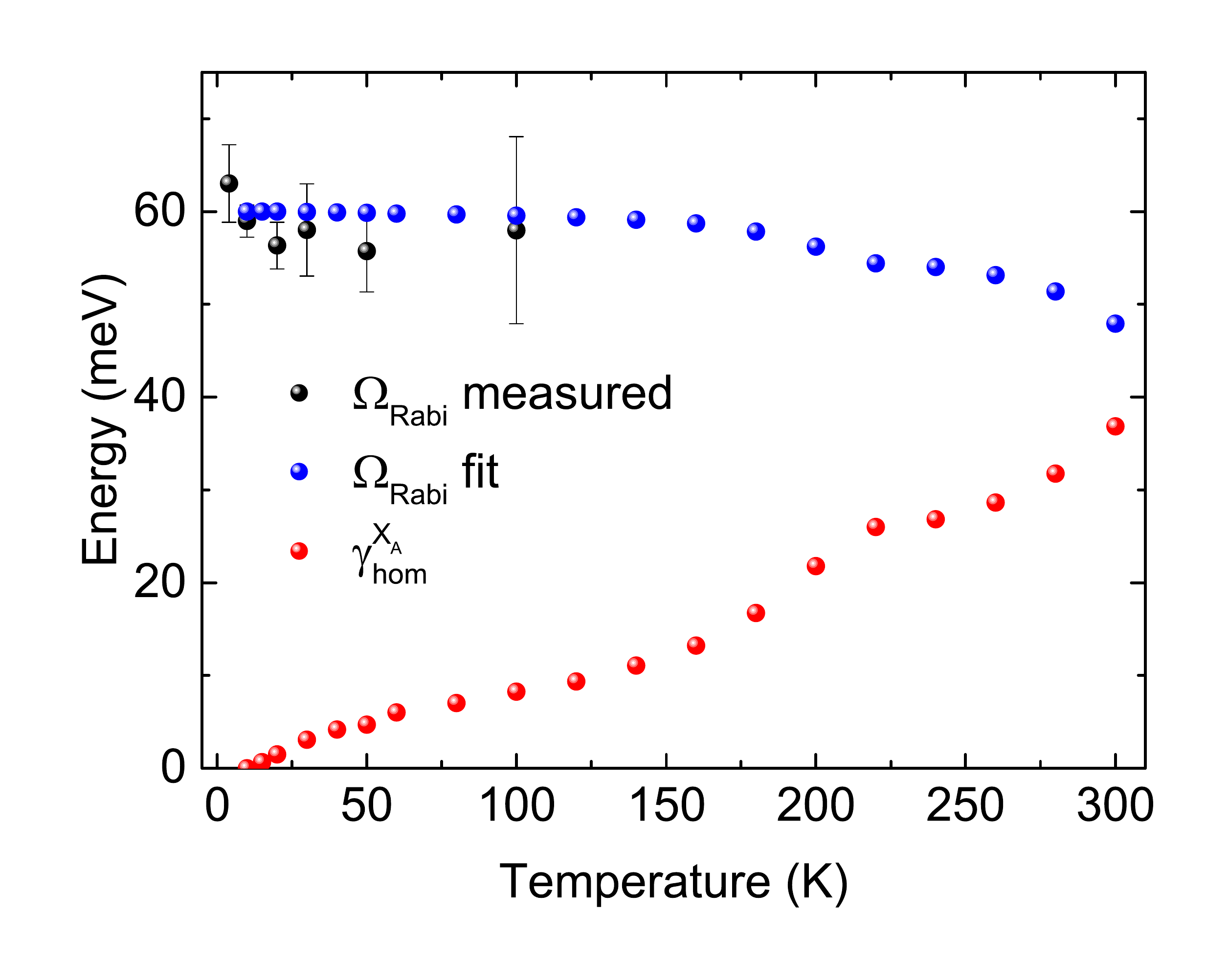}
\caption{Temperature dependence of $\Omega_{Rabi}$. Measured values (black dots) and extrapolation up to room temperature (blue dots) from Eq. \ref{omega}, calculated using the homogeneous X$_A$ exciton linewidth obtained from the measurements shown in Fig. \ref{PL}(a) (red dots).}
\label{VRSvsT}
\end{figure}

Additional information on the temperature dependence of the SCR can be obtained by considering the relation

\begin{equation}
\Omega_{Rabi}(T)=\sqrt{4g_0^2-(\gamma^{X_{A}}_{hom}(T) - \gamma_P)^2},
\label{omega}
\end{equation}

which is derived from the COM.{\cite{suppl}} We assume the photonic linewidth $\gamma_P$ to be temperature-independent. The homogeneous excitonic linewidth of the X$_{A}$ exciton, $\gamma^{X_{A}}_{hom}(T)$, was extracted by fitting a Voigt peak profile to the temperature-dependent PL data shown in Fig. {\ref{PL}}(a). Note that we only take the homogeneous broadening into account, since the normal mode splitting does not depend on the inhomogeneous broadening when the normal mode splitting to MQW inhomogeneous broadening ratio is large.{\cite{Houdre1996,Christmann2006}} The calculated values of $\Omega_{Rabi}(T)$ are shown in Fig. {\ref{VRSvsT}}, together with $\gamma^{X_{A}}_{hom}(T)$. Based on these calculations, we can predict a slight decrease in the normal mode splitting down to 48 meV at 300 K. This again confirms the strong potential of the present platform to investigate the SCR up to room temperature.

\section{Conclusion and Outlook}
\label{conclusion}
In conclusion, we demonstrated propagating polaritons in III-nitride slab waveguides, where \textit{c}-plane GaN/AlGaN MQW excitons hybridize with the propagating TE$_0$ optical mode. The fully lattice-matched growth leads to a very low disorder and high in-plane homogeneity. The SCR was observed up to 100 K with an average normal mode splitting as high as 60 meV due to the large overlap between the QWs and the waveguide mode. The guided polaritons feature a decay length of 50 to 100 $\mu$m and a lifetime of 1-2 ps, which are well accounted for by residual absorption in the QWs.\\

Such a structure shows great potential for photonic and polaritonic integrated circuits up to room temperature. In this perspective, ridge waveguides with lateral confinement would prove a more practical geometry for devices, as it would prevent the isotropic propagation of polaritons. Consequently, the signature of waveguided polaritons should be measurable over longer distances and up to higher temperatures. It would also open the possibility to explore polariton nonlinearities, e.g. polariton soliton wavepackets up to room temperature.

\begin{acknowledgments}
We would like to thank the Swiss National Science Foundation for financial support through Grant No. 200020{\_}162657. 
\end{acknowledgments}


\bibliography{papers}

\begin{thebibliography}{48}
\bibitem [{XBn()}]{XBnote}%
  \BibitemOpen
  \href@noop {} {}\bibinfo {note} {Since we do not observe any spectral
  signature in the polariton dispersion data that could be related to the X$_B$
  exciton, such as the presence of a middle polariton branch, we fully
  neglected X$_B$ in the analysis and the modeling of the dispersion
  curves.}\BibitemShut {Stop}%
\end{thebibliography}

\end{document}


\newcommand{\beginsupplement}{%
        \setcounter{table}{0}
        \renewcommand{\thetable}{S\arabic{table}}%
        \setcounter{figure}{0}
        \renewcommand{\thefigure}{S\arabic{figure}}%
     }
     
\beginsupplement



\pacs{}

\title{Supplementary Material: Propagating Polaritons in III-Nitride Slab Waveguides}


\author{J. Ciers}
\email[]{joachim.ciers@epfl.ch}
\author{J. G. Roch}
\altaffiliation{Currently at Nano-Photonics Group, University of Basel, CH-4056 Basel, Switzerland}
\author{J.-F. Carlin}
\author{G. Jacopin}
\author{R. Butt\'e}
\author{N. Grandjean}

\affiliation{Institute of Physics, \'Ecole Polytechnique F\'ed\'erale de Lausanne (EPFL), CH-1015 Lausanne, Switzerland}

\maketitle


\section{Supplementary Note 1. Definition of $N^{QW}_{eff}$ and $L_{eff}$}

In order to correctly evaluate the coupling between the quantum wells (QWs) and the optical mode, every QW should be weighted by the value of the modal field intensity $\varepsilon \vert E \vert^2$ at its location. The effective number of QWs is therefore:

\begin{equation}
N^{QW}_{eff}=\frac{\sum_{y_i} \varepsilon(y_i) \vert E(y_i) \vert^2}{max(\varepsilon \vert E \vert^2)},
\end{equation}

with $y_i$ the QW positions. In this work, we use the following definition of the effective length of the waveguide mode:

\begin{equation}
L_{eff}=2 \frac{\int \varepsilon \vert E \vert^2 dy}{max(\varepsilon \vert E \vert^2)},
\label{Leff}
\end{equation}

which is consistent with the definition commonly used for microcavities.\cite{Savona1995} It is determined as the cavity length required to have an equal one-dimensional mode volume ($V$) if there were no evanescent field in the distributed Bragg reflectors. The $V$-value of this microcavity mode is then equal to

\begin{equation}
V=\frac{1}{2} L_{eff},
\label{V}
\end{equation} 

since the average value of the modal field intensity over the standing wave is $\frac{1}{2}$. Upon substitution of Eq. \ref{Leff} into Eq. \ref{V}, we recover the usual definition of the mode volume.\cite{Fox2006} Using these definitions, the values of $N^{QW}_{eff}$, $L_{eff}$ and the effective refractive index $n_{eff}$ of the guided mode were calculated using Lumerical's two-dimensional finite-difference time-domain (2D-FDTD) solver.\cite{lumerical} The results are given in Tab. \ref{paramtabFDTD}.

\begin{table}
\caption{\label{paramtabFDTD}Parameters calculated from the 2D-FDTD solver.}
\begin{ruledtabular}
\begin{tabular}{l c r}
Parameter & Value \\
\hline
$L_{eff}$ & 175.6 nm \\
$N^{QW}_{eff}$ & 14.62 \\
$n_{eff}$ & 2.655 \\

\end{tabular}
\end{ruledtabular}
\end{table}

\begin{figure}[h]
\includegraphics[width=0.70\textwidth]{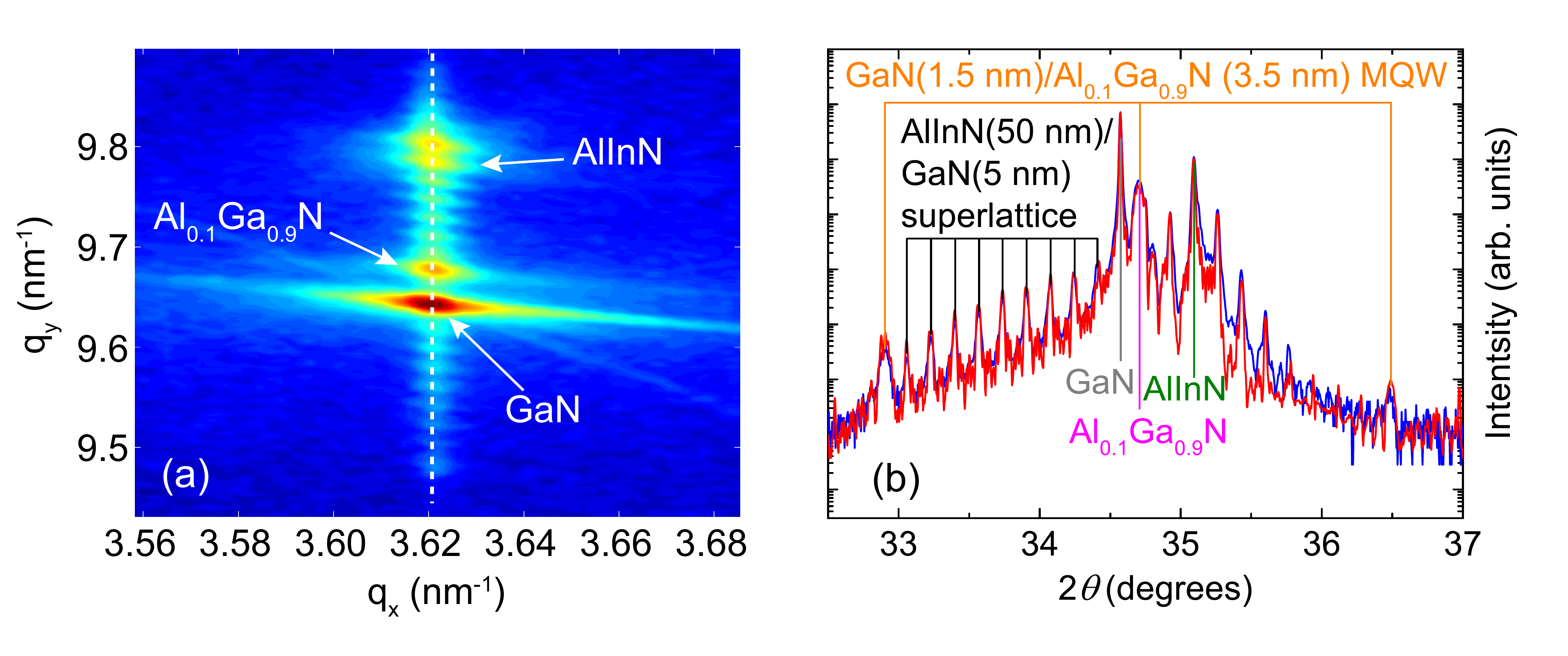}
\caption{X-ray diffraction analysis of the sample. (a) Reciprocal space map of the (10$\bar{1}$5) asymmetric reflections showing that the bottom cladding and the active region are pseudomorphic to the FS-GaN substrate. (b) Coupled scan of the (0002) reflex along the [0001] direction. A model of the structure (red) was fitted to the measured data (blue) with the layer thicknesses and the alloy composition as fitting parameters. The results correspond to the values given in the main text.}
\label{fig2}
\end{figure}

\begin{table}
\caption{\label{paramtab3}Parameters used for the $k \cdot p$ calculations}
\begin{ruledtabular}
\begin{tabular}{l r}
Parameter & Value ($m_0$ is the free electron mass) \\
\hline
Effective electron mass $m^*_e$ (GaN) & $0.2m_0$ \\
Effective hole mass $m^*_h$ (GaN) & $1.4m_0$ \\
Effective electron mass $m^*_e$ (AlGaN) & $0.21m_0$ \\
Effective hole mass $m^*_h$ (AlGaN) & $1.61m_0$ \\
GaN layer thickness & 1.5 nm \\
AlGaN layer thickness & 3.5 nm \\
GaN built-in electric field & -625 kV/cm \\
AlGaN built-in electric field & 220 kV/cm \\

\end{tabular}
\end{ruledtabular}
\end{table}

\section{Supplementary Note 2. Experimental methods}

\begin{figure}{}
\includegraphics[width=0.50\textwidth]{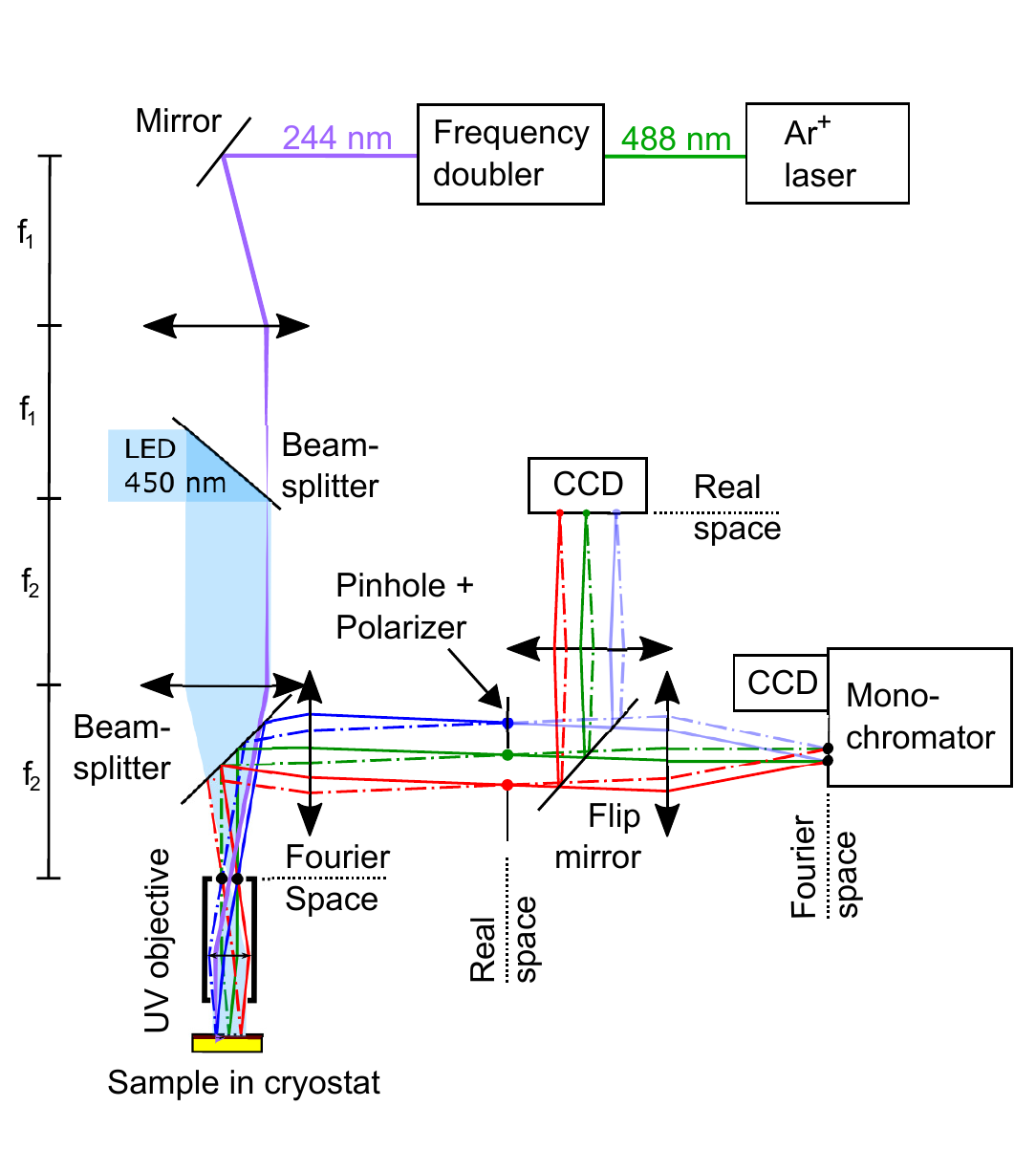}
\caption{Sketch of the Fourier-space spectroscopy set-up. The excitation laser is represented in purple. The light emitted from the sample is represented in red, green and blue with the different colors indicating different locations on the sample. The solid and the dashed lines represent two different emission angles. The Fourier space is marked by black dots and the real space by colored dots.}
\label{figsetup}
\end{figure}

In order to experimentally verify the SCR between the guided photons and QW excitons, the dispersion of the guided modes was measured by Fourier-space spectroscopy using real-space filtering. The experimental set-up is sketched in Fig. \ref{figsetup}. A continuous wave (cw) frequency-doubled Ar$^{+}$ laser emitting at 244 nm was used for the excitation. This laser beam was coupled into a 80$\times$ UV microscope objective -- with 0.55 numerical aperture and a 350 $\mu$m field of view -- through a so-called 4$f$-configuration of two lenses and a mirror. The laser spot diameter on the sample was about 5 $\mu$m. This configuration allowed to scan the excitation spot over the sample within the field of view of the objective in the two dimensions by rotating the mirror. The light emitted by the sample was then collected through the same objective lens and the back focal plane (Fourier plane) was imaged onto the spectrometer entrance slit by two lenses. The spectrometer consists of a liquid-nitrogen cooled back-illuminated UV-enhanced charge-coupled device (CCD) mounted on a 55 cm focal length monochromator. A pinhole was placed in the first real-space image plane of the sample to select only the light emitted from the grating outcoupler. The pinhole diameter of 1.5 mm corresponded to a 19 $\mu$m circle on the sample with the lens combination used here. A wire grid polarizer with an extinction coefficient better than 100:1 between 300 and 400 nm was positioned immediately after the pinhole to perform polarization-dependent measurements. The first real-space image plane was imaged onto a second UV-enhanced CCD for real-space observation of the sample. A blue light-emitting diode (LED) was used for illumination. The sample was mounted in a cold-finger continuous-flow liquid-helium cryostat.\\

Micro-photoluminescence ($\mu$-PL) spectra inside the air light cone were taken using the same setup, without the 4$f$ lenses, spatial filtering pinhole and polarizer. We used an excitation power density of 10 kW/cm$^2$ in an unprocessed area of the sample, i.e. away from any grating. It was verified that no screening of the built-in electric field occurs at these power densities in our MQW sample. Peak shape and position are identical to measurements perfored at much lower power densities (down to 20 mW/cm$^2$).\\

Photoreflectance (PR) measurements near normal incidence were performed with the unprocessed sample in a closed-cycle helium cryostat. Carriers were generated by a frequency-quadrupled Nd:yttrium aluminum garnet (YAG) laser emitting at 266 nm, mechanically chopped at 170 Hz. The laser spot was defocused (diameter of $\sim$1 mm) to ensure a low carrier injection regime and to avoid screening of the built-in electric field in the QWs. The incident probe light on the sample (spot size $\sim$500 $\mu$m) was coming from a xenon lamp coupled to a monochromator. The reflected light intensity was measured by a UV-enhanced Si photodiode and the modulation -- caused by the photogenerated carriers -- was measured by a lock-in amplifier. \\

\begin{figure}{}
\includegraphics[width=0.70\textwidth]{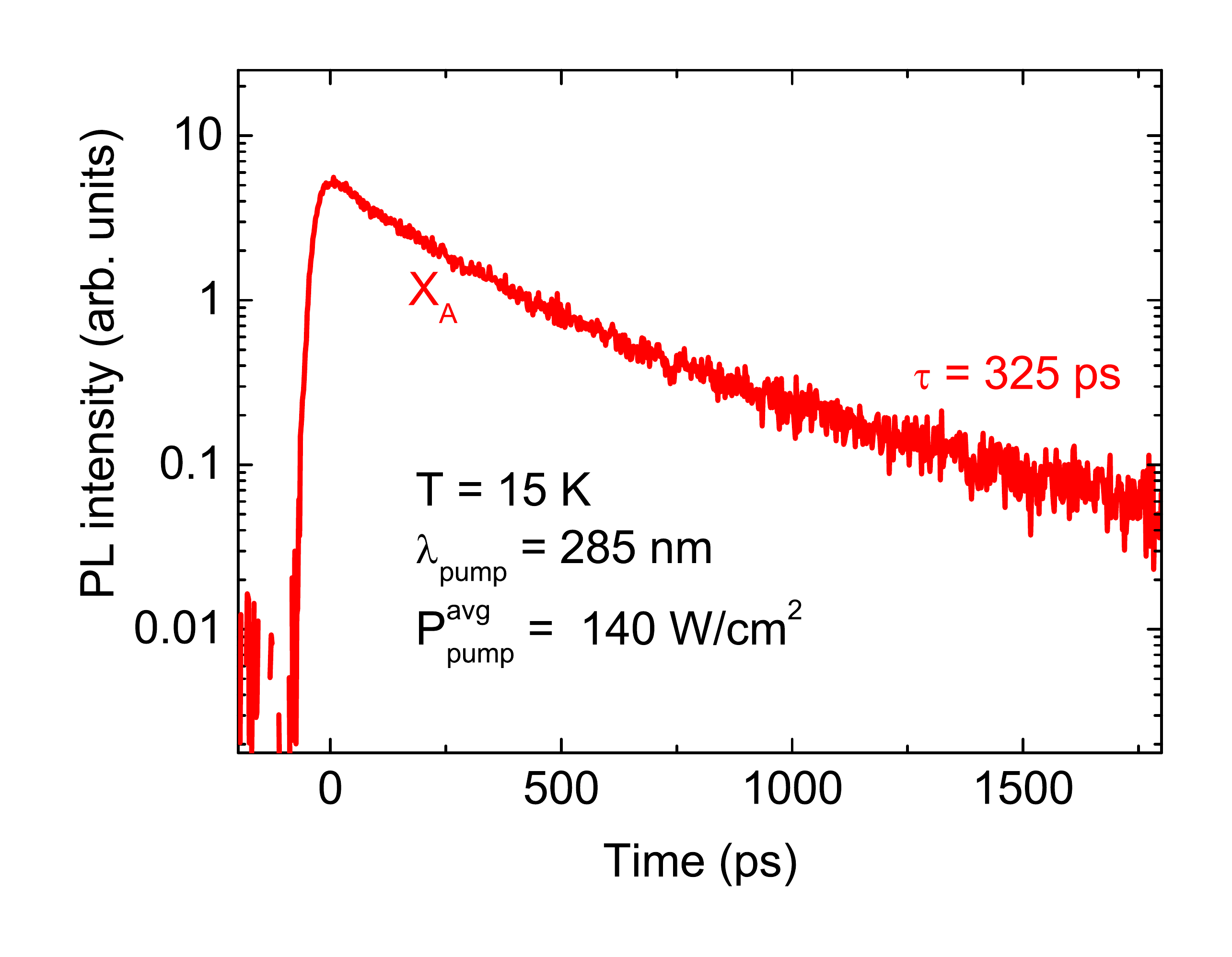}
\caption{TRPL measurement of the X$_A$ exciton transiton at $T$ = 15 K. The short effective lifetime measured here is fully consistent with the marginal impact of the quantum-confined Stark effect in thin QWs.}
\label{TRPL}
\end{figure}

Time-resolved PL (TRPL) experiments were performed with the unprocessed sample in a closed-cycle helium cryostat at $T$ = 15 K, using the third harmonic of a Ti:Al$_2$O$_3$ mode-locked laser (with pulse width of 2 ps and repetition rate of 80.7 MHz). The excitation wavelength was set to 285 nm. We used a sufficiently high average power density of 140 W/cm$^2$ (840 kW/cm$^2$ peak) in order to saturate alternative relaxation channels and determine the radiative recombination lifetime of the free X$_A$ transition. We confirmed that, given the already high electron-hole overlap of 0.8 at zero carrier density, no screening of the built-in electric field occurred for these pump powers. The PL decay was recorded with a Hamamatsu streak camera working in synchroscan mode mounted on a 32 cm focal length monochromator. The results are shown in Fig. \ref{TRPL}.

\begin{figure*}
\includegraphics[width=0.9\textwidth]{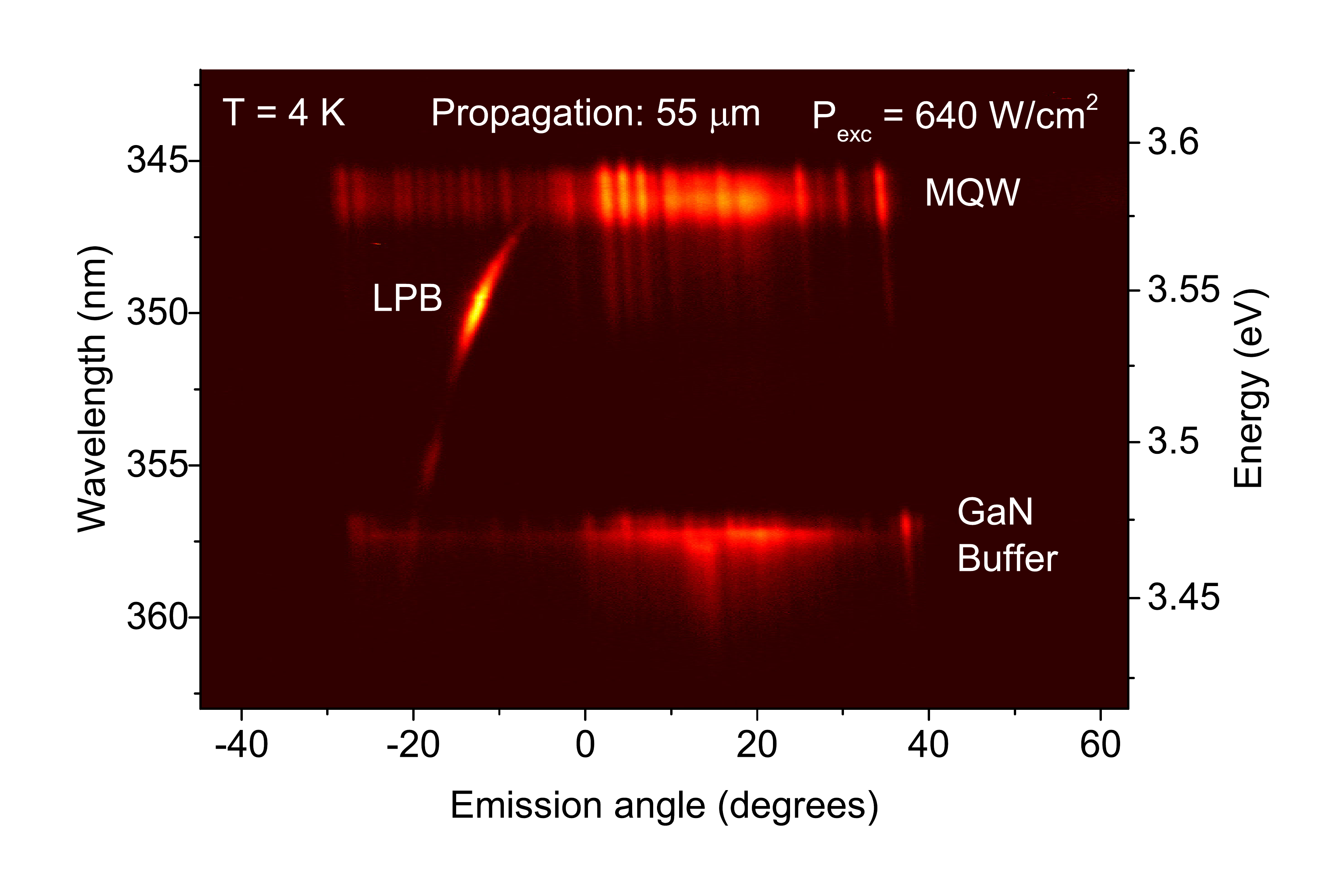}
\caption{Raw angle-resolved photoluminescence (PL) measurement data including the signal from the weakly coupled excitons for the measurement corresponding to Fig. 4 in the main text. We can clearly see the QW and substrate PL signal, together with the LPB signature. Note that the QW emission seems to occur at a lower energy than observed in Fig. 3 in the main text. This is most likely due to the guided character ---due to reflection at the interfaces, but not total internal reflection--- of this signal. The high-energy part of the QW emission is reduced due to absorption near the QW band edge. }
\label{suppdisp}
\end{figure*}

\section{Supplementary Note 3. Finite-difference time-domain simulation of the waveguide modes' dispersion}

To track the resonant modes of the waveguides, 2D-FDTD calculations 
were performed using Lumerical FDTD solutions\citep{lumerical} in the following manner: A broadband plane wave, covering the spectral region of interest, with fixed in-plane wavevector is injected into the waveguide. A power monitor measures the power outcoupled by the grating in the far field as a function of emission angle and wavelength. This power is proportional to the local density of optical states (LDOS) in the waveguide\cite{Taflove2013} and the presence of a peak in the LDOS indicates a resonant mode. We can determine the emission angle and energy of the guided mode corresponding to this particular in-plane wavevector from the peak position. Note that the peak in the LDOS disappears when approaching the exciton energy due to increasing absorption. We limited the spectral range of the calculation at the MQW band edge energy (3.64 eV) on the high-energy side in order to avoid problems related to the cusp shape in the MQW refractive index dispersion close to the band edge. A sweep over the in-plane wavevectors then gives access to the full mode dispersion. The parameters used to model the refractive index of the layers are summarized in Table \ref{paramtab1}. $E_X$, $E_B$, $\Gamma_{hom}$ and $\Gamma_{inh}$ are the exciton energy, exciton binding energy, homogeneous and inhomogeneous broadening of the exciton transition, respectively.

\begin{table}
\caption{\label{paramtab1}Parameters used for the 2D-FDTD calculations at $T$ = 4 K.}
\begin{ruledtabular}
\begin{tabular}{l c r}
Parameter & Value at 4 K & Comment \\
\hline
$E_X$(AlGaN) & 3.683 eV & Deduced from PR measurements\\
$E_B$(AlGaN) & 28.1 meV & Linear interpolation between bulk values of GaN (25 meV)\cite{Kornitzer1999} and AlN (56 meV)\cite{Rossbach2011a}\\
$E_X$(QW) & 3.599 eV & Deduced from PR measurements\\
$E_B$(QW) & 40 meV\cite{Leavitt1990} & \\
$\Gamma_{inh}$(QW) & 8 meV & Deduced from PL measurements\\
$\Gamma_{hom}$(QW) & Negligible & \\

\end{tabular}
\end{ruledtabular}
\end{table}

\begin{figure*}
\includegraphics[width=0.9\textwidth]{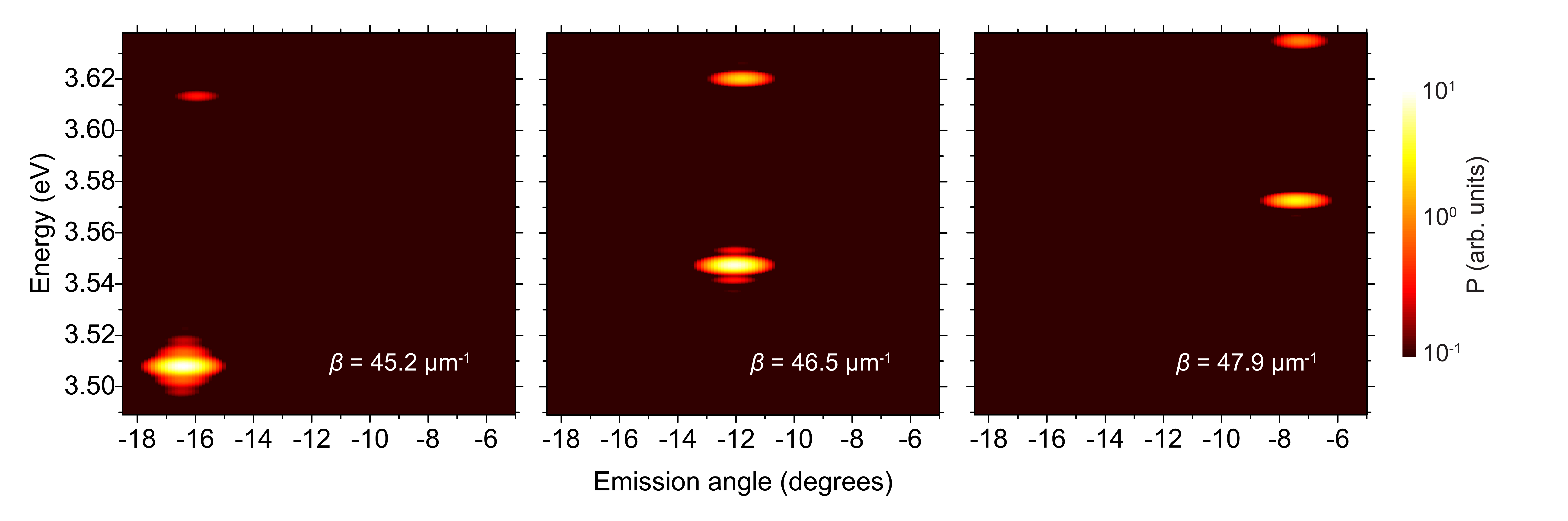}
\caption{Illustration of the 2D-FDTD calculations of the waveguide mode dispersion. The emitted power $P$ from the waveguide grating is calculated in the far field as a function of emission angle and energy for a broadband source with fixed in-plane wavevector $\beta$. The peak in emitted power corresponds to the guided mode spectral position. Calculations are repeated for different values of $\beta$ to reproduce the modal dispersion. }
\label{supplFDTD}
\end{figure*}

\section{Supplementary Note 4. Variable stripe length measurements}
\begin{figure*}
\includegraphics[width=0.7\textwidth]{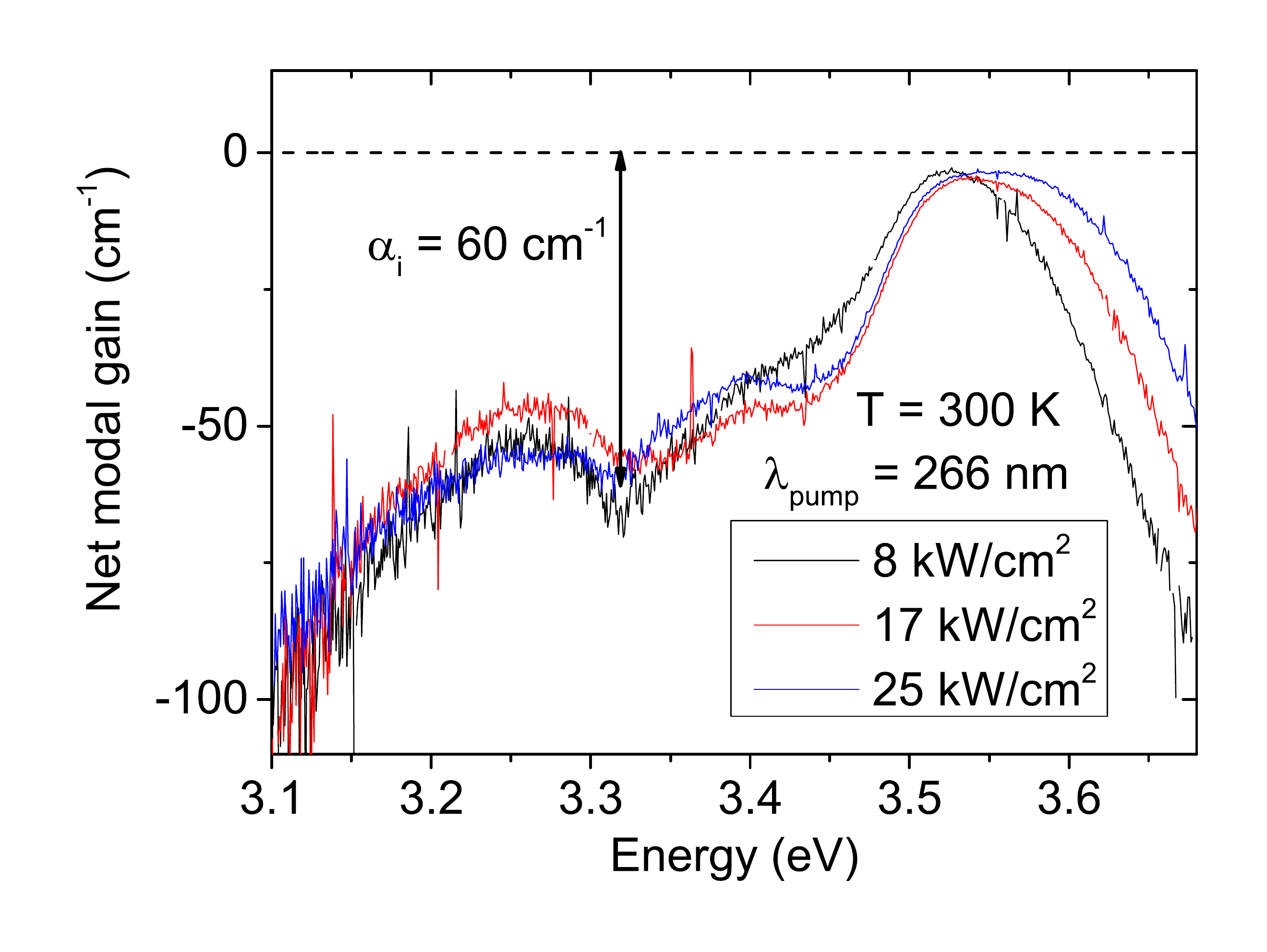}
\caption{VSL measurements of the present waveguide leading to an intrinsic absorption of 60 cm$^{-1}$ below the gain band.}
\label{suppVSL}
\end{figure*}

Variable stripe length (VSL) measurements were performed on the waveguide sample at 300 K to determine residual absorption losses from the net modal gain. \cite{Shaklee1973} The sample is pumped from the sample surface with a 25 $\mu$m wide and a 400 $\mu$m ($L$) or 800 $\mu$m ($2L$) long rectangular laser spot. We use a 266 nm frequency-quadrupled Nd:YAG laser, emitting 440 ps long pulses with a repetition rate of 9 kHz, which is expanded using a beam expander and then refocused with a cylindrical lens. The light generated in the waveguide is collected from a cleaved facet with an optical fiber in the end-fire geometry and coupled to a spectrometer.  We can determine the net modal gain $G$ by comparing the intensities of the measured signal for the two stripe lengths $I_{2L}$ and $I_{L}$:

\begin{equation}
G(\lambda) = \Gamma g(\lambda) -\alpha(\lambda) = \frac{1}{L}\ln \left(\frac{I_{2L}(\lambda)}{I_{L}(\lambda)} -1 \right),
\end{equation}

where  $\Gamma$, $g(\lambda)$, and $\alpha(\lambda)$ are the optical confinement factor, material gain, and absorption coefficient, respectively. The resulting spectra are shown in Fig. \ref{suppVSL}. We can determine the intrinsic absorption $\alpha_i$ from these spectra below the gain band, and find a value of 60 cm$^{-1}$ at around 3.32 eV, which is 200 meV below the $X_A$ energy at room temperature (3.524 eV). We assume this value to be constant with temperature, for a constant relative energy to $X_A$. This approximation is valid as long as the energy separation to the exciton resonance is substantially larger than the broadening of the transition.

\section{Supplementary Note 5. Coupled Oscillator Model}

In the coupled oscillator model, the coupled photonic and excitonic states are described by the Hamiltonian in matrix form

\begin{equation}
\widehat{H} = 
\begin{bmatrix}
    E_X + i\gamma_X       & g_0 \\
    g_0       & E_P + i\gamma_P  \\    
\end{bmatrix},
\end{equation}

where $E_X$, $\gamma_X$ and $E_P$, $\gamma_P$ are the energy and decay rate of the exciton and the photon, respectively, and $g_0$ is the coupling strength between both modes. The eigenstates of the system are the upper (UPB) and lower polariton branch (LPB) and can be found by diagonalizing this Hamiltonian:

\begin{equation}
E_{UPB}=\frac{1}{2}\left[ E_X + E_P + i(\gamma_X + \gamma_P) + \sqrt{4g_0^2 - (\gamma_X - \gamma_P)^2 +  (E_X - E_P)^2   }\right] 
\end{equation}
\begin{equation}
E_{LPB}=\frac{1}{2}\left[ E_X + E_P + i(\gamma_X + \gamma_P) - \sqrt{4g_0^2 - (\gamma_X - \gamma_P)^2 +  (E_X - E_P)^2   }\right]
\label{LPB} 
\end{equation}

The normal mode splitting $\Omega_{Rabi}$ is defined at zero detuning ($E_X = E_P$). By subtracting the two energies, we obtain

\begin{equation}
\Omega_{Rabi}=\sqrt{4g_0^2-(\gamma_X - \gamma_P)^2}.
\end{equation}

Eq. \ref{LPB} was fitted to the measured dispersion curves with $g_0$ as a free parameter.

\bibliography{papers}